\documentclass[epjCONF]{svjour}
\usepackage{graphicx}
\usepackage[varg]{txfonts} 
\usepackage[latin1]{inputenc}
\usepackage{epsfig}
\usepackage{epstopdf}
\usepackage{color}
\usepackage{xspace}
\newcommand{\stwo}{\mbox{\ensuremath{\sqrt{s_{\mathrm{NN}}}=2.76\;\mathrm{TeV}}}\xspace}
\newcommand{\ssvn}{\mbox{\ensuremath{\sqrt{s}=7\;\mathrm{TeV}}}\xspace}

\session-title{Resonance Workshop at UT Austin}
\begin{document}
\title{Hadronic Resonances in Heavy-Ion Collisions at ALICE}
\author{A. G. Knospe\thanks{\email{anders.knospe@cern.ch}} for the ALICE Collaboration}
\institute{The University of Texas at Austin, Austin, Texas, USA}
\abstract{
Modifications to the masses and widths of hadronic resonances in heavy-ion collisions could be a sign of chiral symmetry restoration.  Uncorrected spectra, masses, and widths of the $\phi(1020)$ and $\mathrm{K}^{*}(892)^{0}$ resonances have been measured in Pb--Pb collisions at \stwo using the ALICE detector.  These measurements are presented and compared to resonances in other collision systems.
} 
\maketitle
\section{Introduction}
\label{sec:intro}

Hadronic resonances are important probes of heavy-ion collisions and can allow us to study the properties of the system at different stages of its evolution.  Resonances have short enough lifetimes (a few fm/$c$) that they may decay before freeze-out.  The products of their hadronic decays may be re-scattered due to interactions in the hadronic medium~\cite{Bleicher_Aichelin,Bleicher_Stoecker,Markert_thermal,Vogel_Bleicher}, reducing the measured resonance signal\footnote{Measurements of resonance yields through leptonic decay channels are beyond the scope of these proceedings.}.  Resonances may also be regenerated in the hadronic medium through pseudo-elastic collisions of hadrons~\cite{Bleicher_Aichelin}.  The competition between resonance generating processes and re-scattering, and therefore the ratio of resonance yields to non-resonance yields, is governed by the lifetime and temperature of the hadronic medium.  Thermal models \cite{Markert_thermal,Andronic2009,PBM2011,AndronicQM2011,Torrieri_thermal} can be used to predict particle ratios as functions of the chemical freeze-out temperature $(T_{\mathrm{ch}})$ and the time between chemical and thermal freeze-out.  Comparing model predictions to measurements of particle ratios allows us to tune the model parameters and extract estimates of $T_{\mathrm{ch}}$ and the lifetime of the hadronic phase.

The study of resonances can also provide information about chiral symmetry restoration in the partonic and hadronic phases.  If resonances are produced in the QGP\footnote{The phase transition between QGP and hadronic matter is expected \cite{Karsch} at a critical temperature of \mbox{$T_{\mathrm{c}}=173\pm8\;\mathrm{MeV}$}.}, changes in their masses and/or widths may be observed~\cite{Rapp2009,Brodsky_chiral}.  A resonance may be produced off-shell or interact with the medium, leading to a shift in the observed mass.  Interactions with the medium may cause a resonance to dissociate, leading to a reduction in its lifetime, or an increase in its width.  Some models \cite{Holt_Haglin} call for an increase in resonance widths by up to a factor of 10.

These proceedings describe the measurement of hadronic resonances in Pb--Pb collisions at\linebreak \stwo using the ALICE detector, concentrating on measurements of the $\phi(1020)$ and\linebreak$\mathrm{K}^{*}(892)^{0}$, hereinafter referred to as $\phi$ and $\mathrm{K}^{*0}$.  The method used to extract these resonances from heavy-ion collisions is described in Section \ref{sec:analysis:method}.  Uncorrected spectra, and measurements of masses and widths are presented in Section \ref{sec:analysis:results}.  A method of selecting resonances that decayed earlier in the collision, when chiral symmetry was restored, is presented in Section \ref{sec:correlations}.

\section{Analysis of $\phi$ and $\mathrm{K}^{*0}$}
\label{sec:analysis}

\subsection{Analysis Method}
\label{sec:analysis:method}


\begin{figure}
\resizebox{1.0\columnwidth}{!}{
\includegraphics{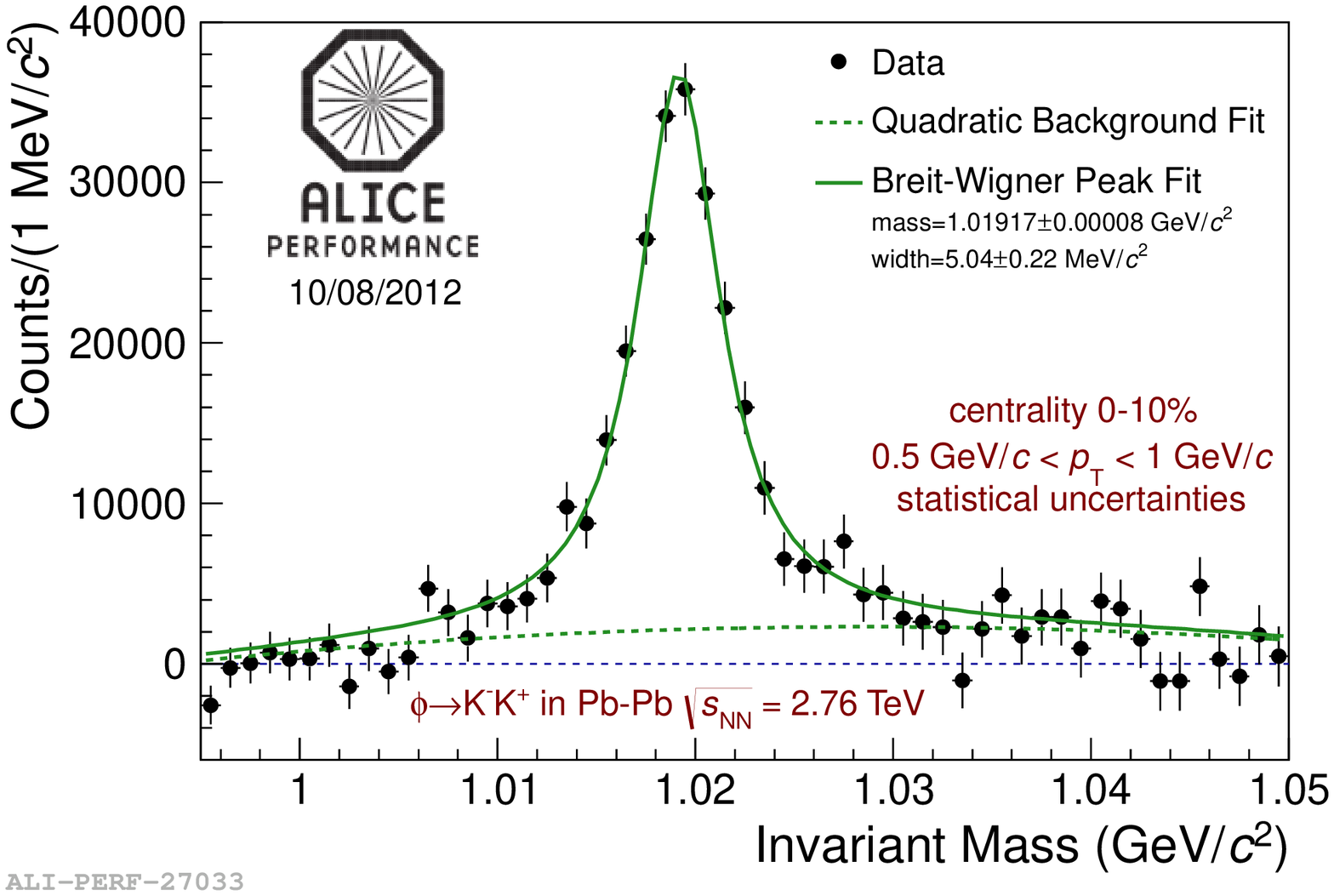}
\includegraphics[scale=0.81]{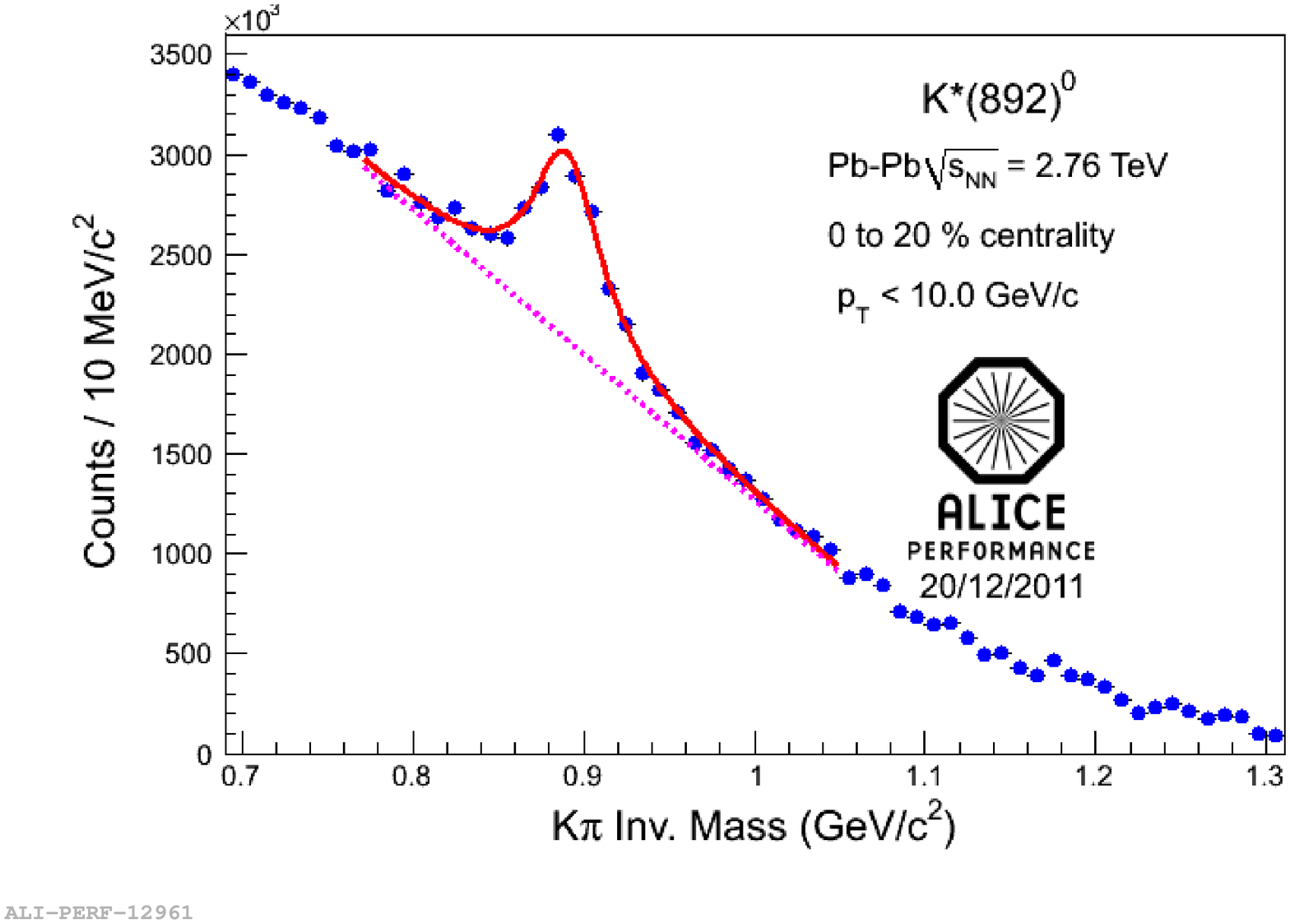} }
\caption{\textbf{Left plot:} Fit of background-subtracted invariant mass distribution of $\phi$ mesons in Pb--Pb collisions at \stwo, centrality 0-10\%, 0.5 GeV/$c < p_{\mathrm{T}} <$ 1 GeV/$c$.  The distribution was generated by subtracting the like-charge combinatorial background (KK pairs with the same charge in the same event) from the distribution of un-like charge KK pairs in the same event.  Dashed curve: quadratic fit excluding peak to describe residual background.  Solid curve: Breit-Wigner peak added to residual background.  \textbf{Right plot:} Fit of background-subtracted invariant mass distribution of $\mathrm{K}^{*0}$ resonance in Pb--Pb collisions at \stwo, centrality 0-20\%, $p_{\mathrm{T}} <$~10~GeV/$c$.  Dashed curve: linear residual background fit.  Solid curve: Breit-Wigner peak added to residual background.}
\label{fig:analysis:invmass_sub}
\end{figure}

The ALICE detector is described in detail in \cite{ALICE_detector}.  This analysis uses the Inner Tracking System (ITS) for tracking\footnote{The ITS is not used for particle identification, but ITS hits are used in track reconstruction.  Information from the ITS is used in the reconstruction of the primary vertex and in the reconstruction of the momentum for each track.}, the Time Projection Chamber (TPC) for tracking and particle identification, and the V0 detector, which provides minimum-bias triggers.  The $\phi$ mesons were identified by reconstructing the decay channel $\phi\rightarrow \mathrm{K}^{-}\mathrm{K}^{+}$ (branching \mbox{ratio = $0.489\pm 0.005$~\cite{PDG}}) in 9.5 million minimum-bias Pb--Pb collisions.  The kaons daughters were identified using their energy loss in the TPC.  The measured energy loss was required to be within $2\sigma_{\mathrm{K}}$ of the mean energy loss for kaons (at the given momentum).  The unit $\sigma_{\mathrm{K}}$ is the expected variance of the kaon energy-loss distribution about its mean value\footnote{This energy loss resolution is less than 5\% for isolated tracks.\cite{ALICE_TPC}}.  The invariant mass of each pair of oppositely charged kaons in the same event was reconstructed; a rapidity cut of $|y_{\mathrm{pair}}|<0.5$ was applied.  The combinatorial background was constructed using two different methods.  First, the like-charge (or ``like-sign") combinatorial background was constructed from pairs of charged kaons (in the same event) with the same charge.  In each invariant-mass bin, the value of the like-charge background is $2\sqrt{n_{--}n_{++}}$, where $n_{--}$ $(n_{++})$ is the number of $\mathrm{K}^{-}\mathrm{K}^{-}$ $(\mathrm{K}^{+}\mathrm{K}^{+})$ pairs in the bin.  Second, the mixed-event combinatorial background was constructed by finding the invariant mass distributions of kaon pairs from separate events.  The two events mixed in this fashion are required to have similar $z$-vertex positions and centralities.  Each kaon track was mixed with tracks from about 5 other events.  For each combination, the difference in the $z$-vertex position was required to be less than 5 cm, while the difference in centrality percentile was required to be less than 10\%.  The mixed event background was normalized so as to yield the same integral as the unlike-charge distribution in a region surrounding but excluding the $\phi$ peak.  Since the choice of the exact boundaries of the normalization region is somewhat arbitrary, six different normalization regions were tried, with different outer and/or inner boundaries\footnote{The normalization region is defined as ($B_{1}\;<\;M_{\mathrm{inv}}(\mathrm{KK})\;<\;B_{2}$ or $B_{3}\;<\;M_{\mathrm{inv}}(\mathrm{KK})\;<\;B_{4}$), with the following sets of boundaries used: $(B_{1},B_{2},B_{3},B_{4})$=(1,1.01,1.03,1.06), (1,1.006,1.04,1.06), (1,1.006,1.03,1.06), (1,1.01,1.035,1.06), (1,1.006,1.04,1.05), and (0.995,1.01,1.03,1.06) (units of GeV/$c^{2}$).}.  The differences in the yield due to the choice of normalization region are typically a few percent and will be incorporated into the systematic uncertainties.  The mixed-event combinatorial background was the primary background used.  The difference in the yield due to the choice of combinatorial background is typically 5-15\% and will be incorporated into the systematic uncertainties.  After subtracting the combinatorial background the $\phi$ peak sits on top of a correlated residual background (usually small).  This residual background is parametrized by fitting the invariant mass distribution with a quadratic polynomial in a region surrounding but excluding the peak (see Figure \ref{fig:analysis:invmass_sub}).  Linear and cubic polynomials have also been tried; differences in the final results due to these alternate fitting functions are incorporated into the systematic uncertainties.  The yield of $\phi$ mesons is extracted by integrating the unlike-charge invariant mass distribution and subtracting off the integral of the residual background.  The unlike-charge invariant mass distribution is also fit with a combined function: the residual background polynomial plus a relativistic Breit-Wigner function to describe the peak.  A Voigtian peak (the convolution of a Breit-Wigner function and a Gaussian to account for momentum resolution) will be tried in the future.  The mass and width of the $\phi$ meson peak are extracted from the peak fit.

The $\mathrm{K}^{*0}$ mesons were identified by reconstructing the decay channel $\mathrm{K}^{*0}\rightarrow\pi^{\pm}\mathrm{K}^{\mp}$ (branching \mbox{ratio = $0.66601\pm0.00006$~\cite{PDG}}) in 3.4 million minimum-bias Pb--Pb collisions.  The decay daughters were identified using their energy loss in the TPC.  A cut of $\pm 2\sigma_{\pi}$ $(\pm 2\sigma_{\mathrm{K}})$ was applied to identify the pions (kaons).  The $\mathrm{K}^{*0}$ analysis was conducted in the same manner as the $\phi$ analysis, although in this case a linear fit function was used as the primary function to parametrize the residual background.

\subsection{Results}
\label{sec:analysis:results}

\begin{figure}
\resizebox{1.0\columnwidth}{!}{
\includegraphics{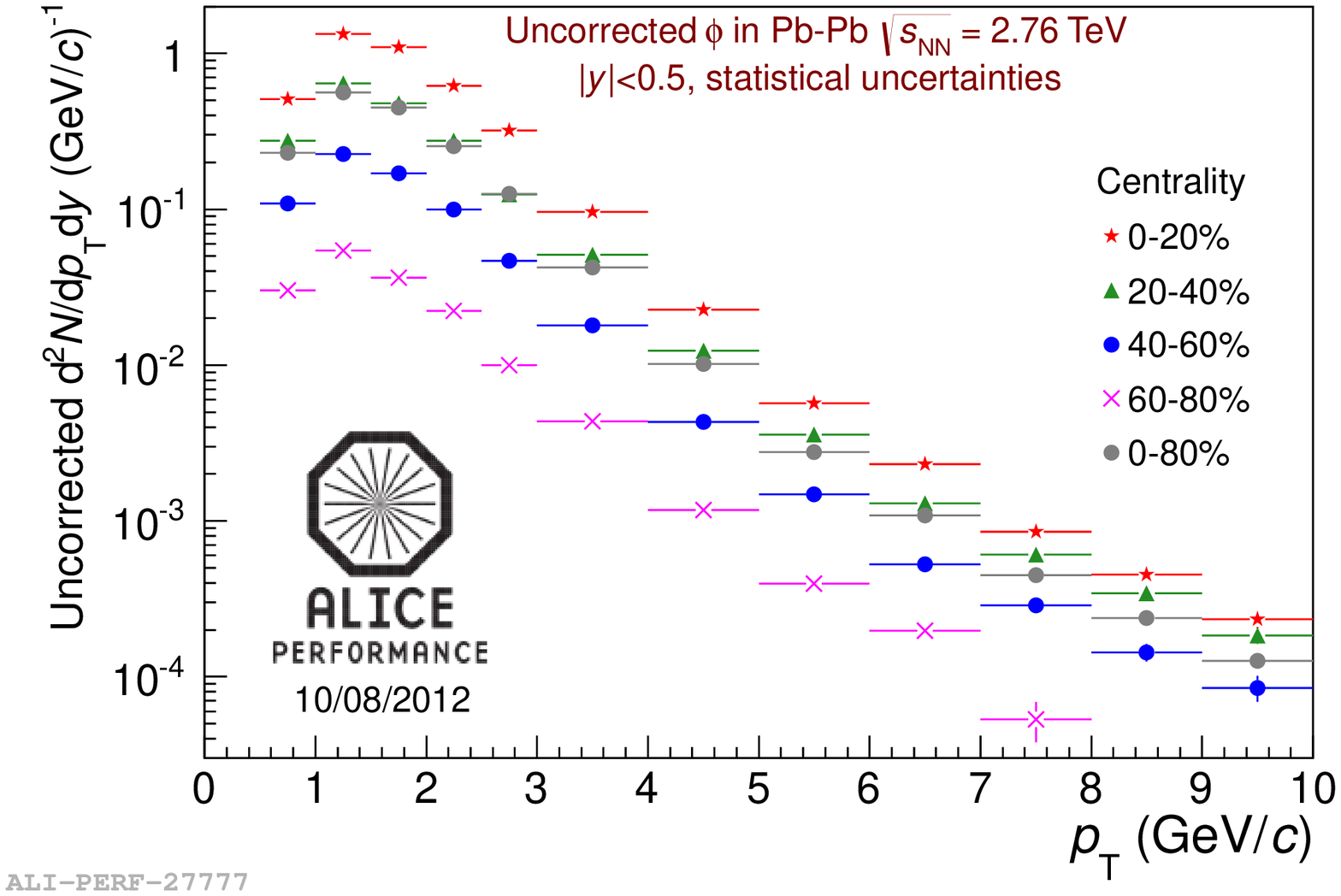}
\includegraphics{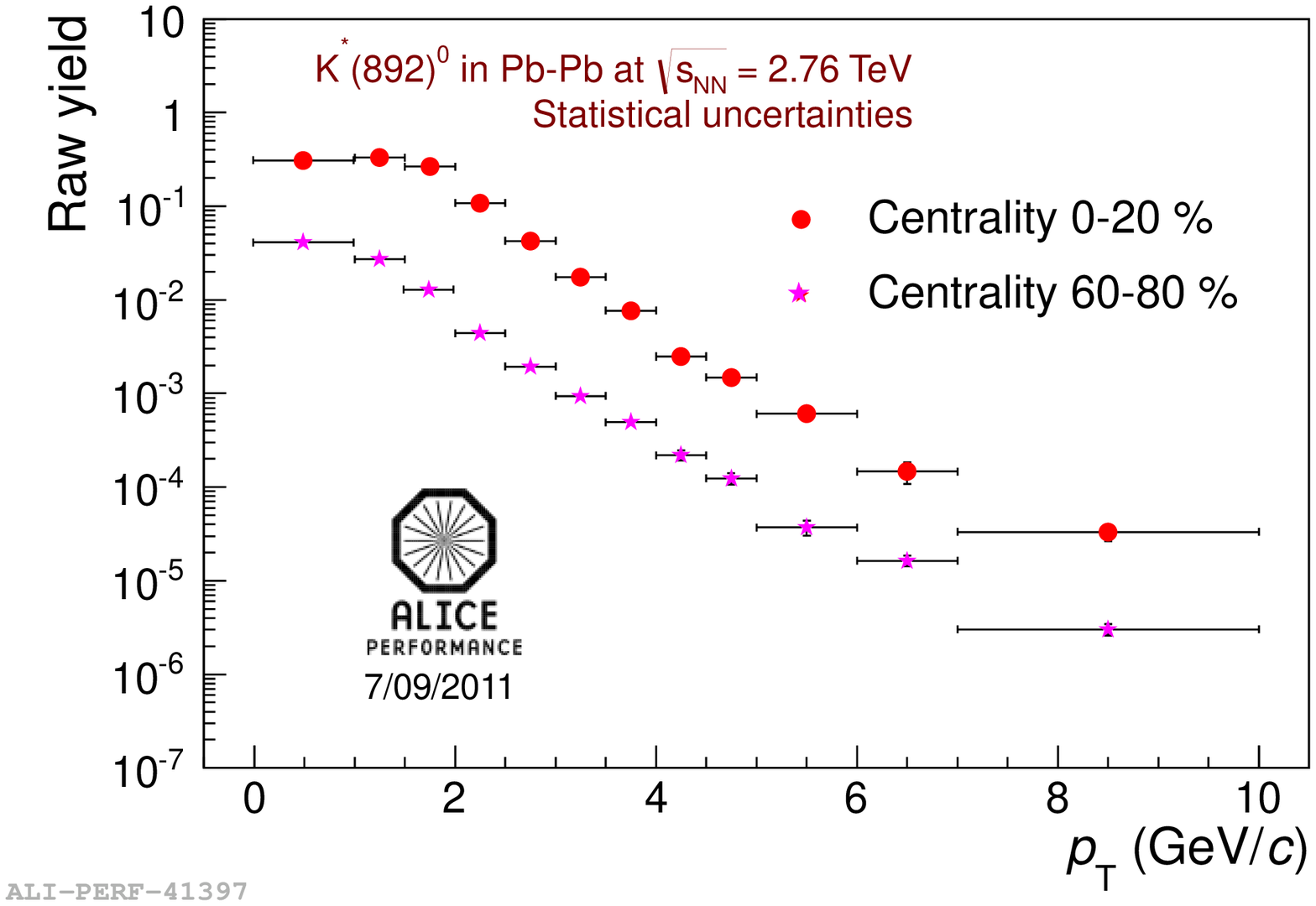} }
\caption{\textbf{Left plot:} Uncorrected yields of $\phi$ mesons in Pb--Pb collisions at \stwo in multiple centrality bins.  The raw counts have been normalized to the number of events, the $\phi\rightarrow \mathrm{K}^{-}\mathrm{K}^{+}$ branching ratio, and the widths of the $p_{\mathrm{T}}$ and rapidity bins $(\mathrm{d}y=1)$.  \textbf{Right plot:} Uncorrected yields of $\mathrm{K}^{*0}$ mesons in Pb--Pb collisions at \stwo in two (central and peripheral) centrality bins.}
\label{fig:analysis:spectra}
\end{figure}


\begin{figure}
\resizebox{1.0\columnwidth}{!}{
\includegraphics{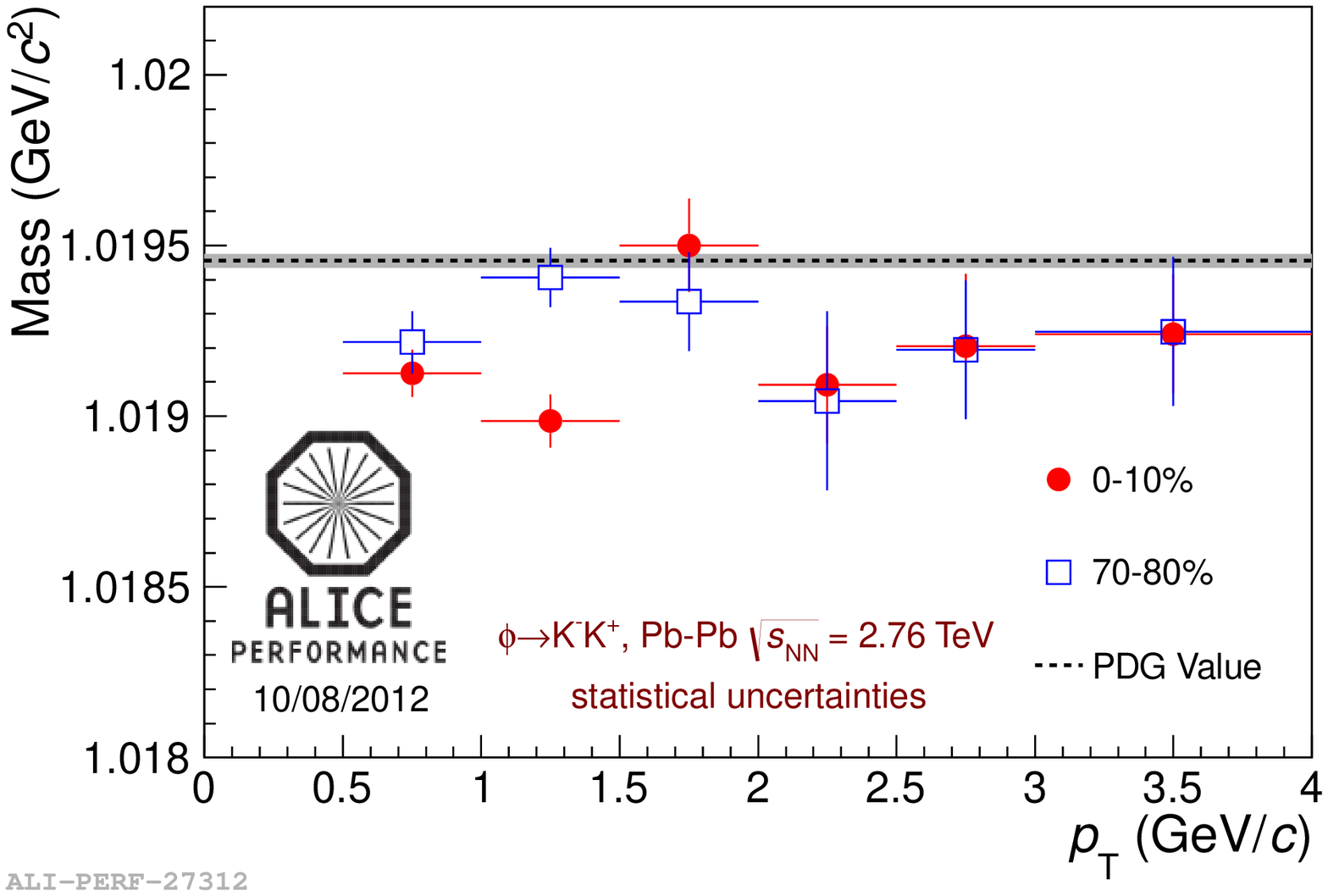}
\includegraphics{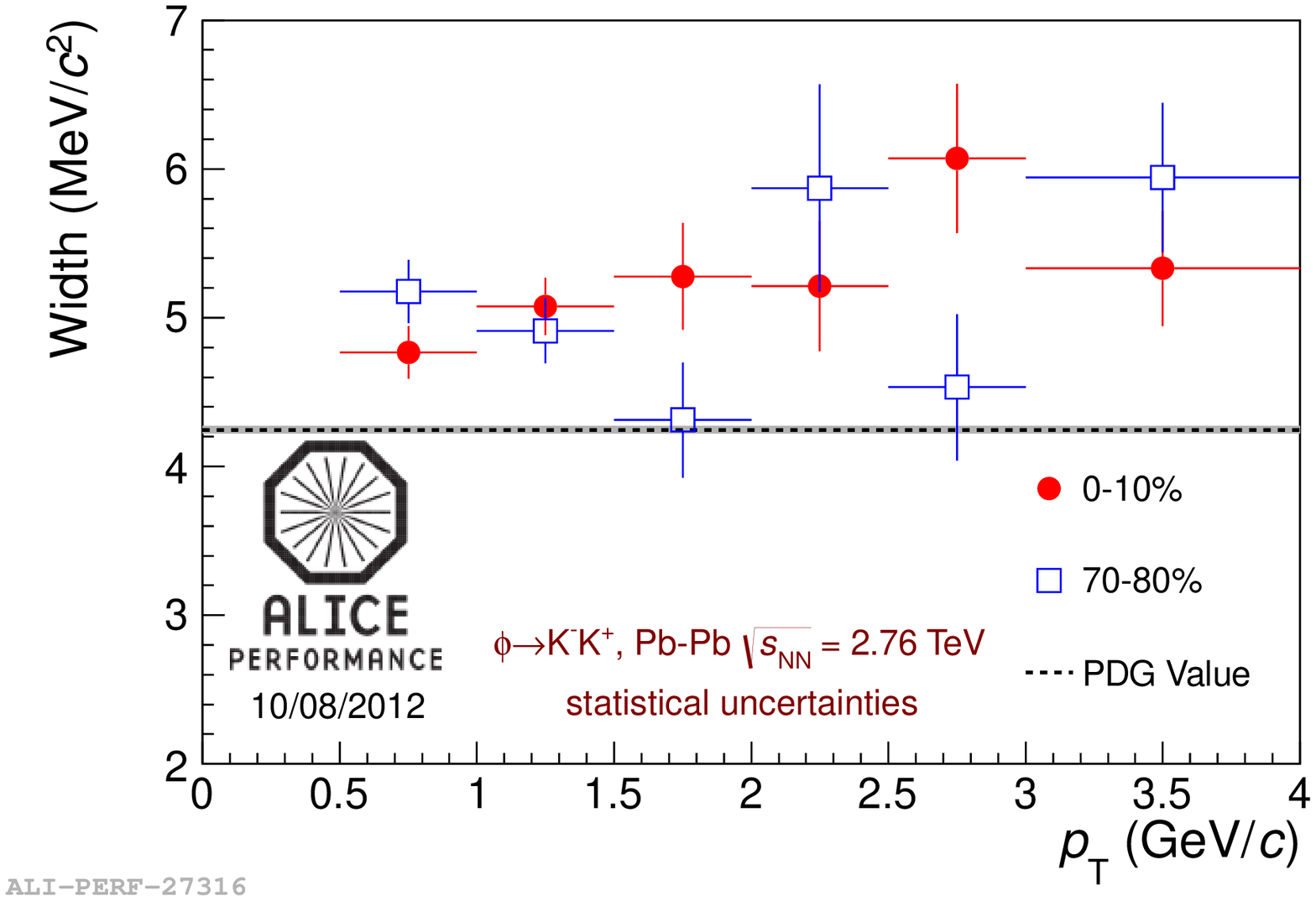} }
\caption{Mass (left plot) and width (right plot) of $\phi$ mesons as a function of $p_{\mathrm{T}}$ for central (0-10\%) and peripheral (70-80\%) Pb--Pb collisions at \stwo.  The vacuum values of the mass and width are shown as dashed lines, with gray bands giving their uncertainties.}
\label{fig:analysis:mass_width_phi}
\end{figure}


\begin{figure}
\resizebox{1.0\columnwidth}{!}{
\includegraphics{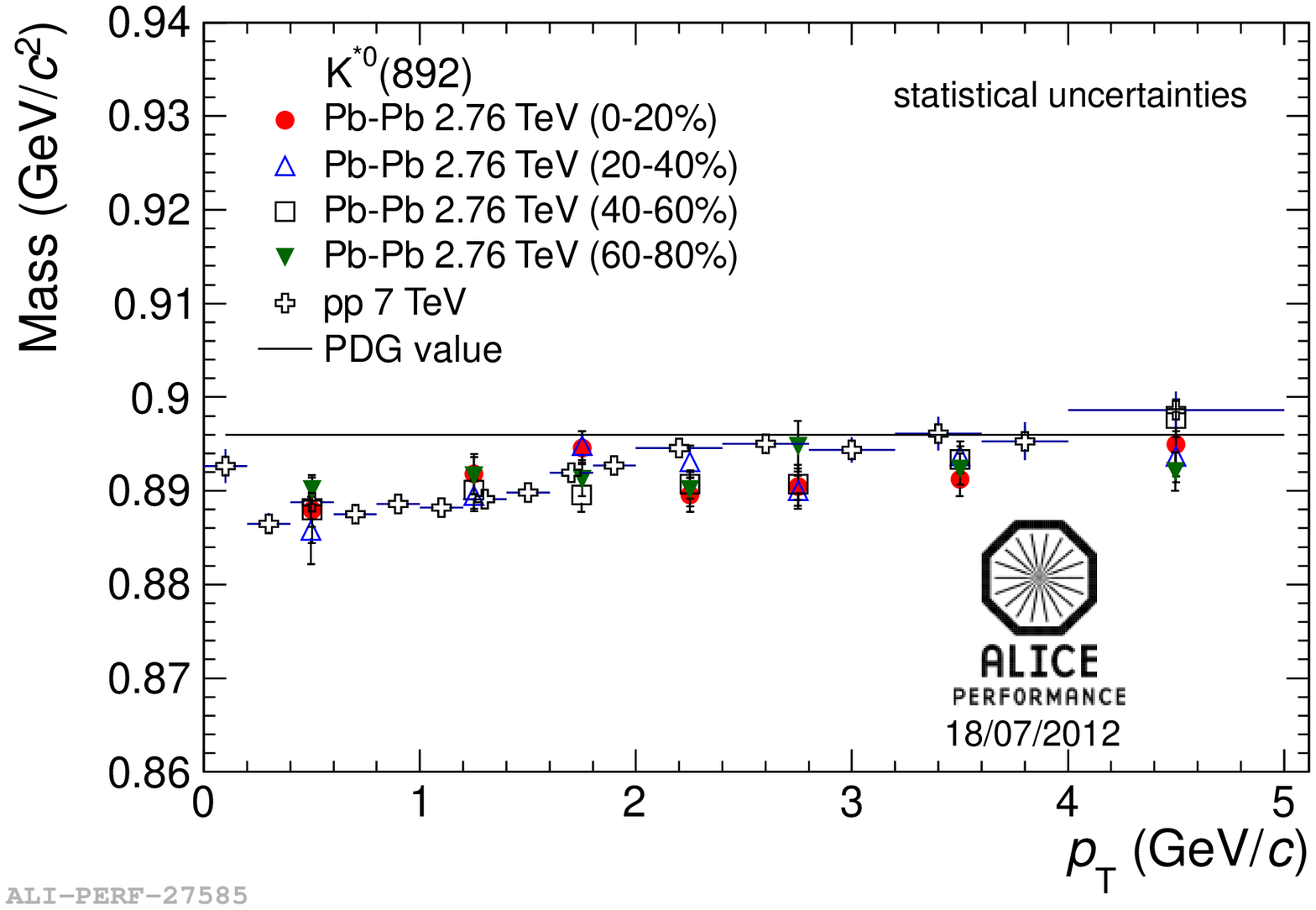}
\includegraphics{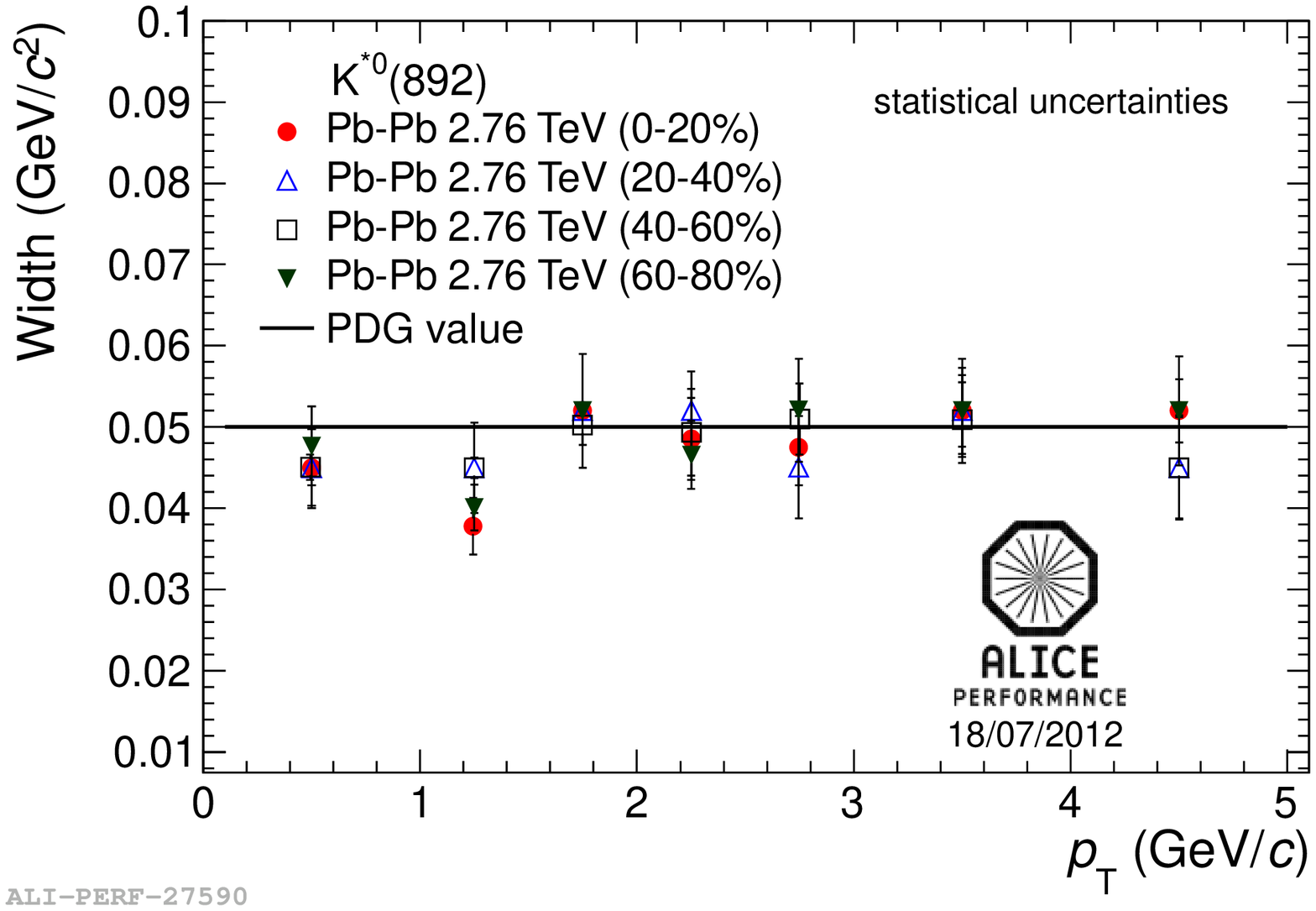} }
\caption{\textbf{Left plot:} Mass of $\mathrm{K}^{*0}$ as a function of $p_{\mathrm{T}}$ in Pb--Pb collisions at \stwo for different centrality bins. Also shown are results for pp collisions at \ssvn.  The black line represents the vacuum value.  Only statistical uncertainties are shown.  \textbf{Right plot:} Width of $\mathrm{K}^{*0}$ as a function of $p_{\mathrm{T}}$ in Pb--Pb collisions at \stwo for different centrality bins.  The black line represents the vacuum value.  Only statistical uncertainties are shown.}
\label{fig:analysis:mass_width_kstar}
\end{figure}


Figure \ref{fig:analysis:spectra} shows the uncorrected $\phi$ and $\mathrm{K}^{*0}$ spectra for Pb--Pb collisions at \stwo.  Resonance signals are observed out to $p_{\mathrm{T}}$=10 GeV/$c$.  These spectra will be corrected for the acceptance, tracking efficiency, and particle identification efficiency.  The corrected spectra will be integrated to find the total yield of these resonances and their ratios to non-resonances (specifically $\pi^{\pm}$ and $\mathrm{K}^{\pm}$), which can be compared to thermal model predictions.

Figure \ref{fig:analysis:mass_width_phi} shows the measured masses and widths of $\phi$ mesons as functions of transverse momentum for three centrality bins (statistical uncertainties only).  No centrality dependence is observed in either quantity.  The $\phi$ mass is found to be compatible with its vacuum value~\cite{PDG} within 0.5 MeV/$c^{2}$; the width is found to be compatible with its vacuum value within 2 MeV/$c^{2}$.  Similar masses and widths are obtained by performing the same fitting procedure on data from Monte-Carlo simulations to account for the momentum resolution and the interactions of the decay daughters with the detector material (particle production from HIJING, particle interactions with the detector simulated using GEANT3).

Figure \ref{fig:analysis:mass_width_kstar} shows the measured masses and widths of $\mathrm{K}^{*0}$ mesons as functions of transverse momentum for multiple centrality bins (statistical uncertainties only).  As observed for the $\phi$ meson, $\mathrm{K}^{*0}$ masses and widths in Pb--Pb collisions do not exhibit any centrality dependence.  The $\mathrm{K}^{*0}$ masses measured in Pb--Pb collisions are consistent with the masses measured in pp collisions at \ssvn~\cite{Markert_SQM2011}, this suggests that the deviation from the vacuum value observed at low $p_{\mathrm{T}}$ is not an effect of the medium.  Detector effects, which may account for the deviation from the vacuum value, are still under investigation.  The $\mathrm{K}^{*0}$ widths are consistent with the the vacuum value.


\section{Hadron-$\phi$ Correlations}
\label{sec:correlations}

The hadron-resonance correlation method is designed to select for resonances that have a larger probability of exhibiting the signatures of chiral symmetry restoration, \textit{i.e.}, resonances that interacted with and decayed inside the QGP.  The method is summarized in Section ~\ref{sec:correlations:method} and results (hadron-$\phi$ correlations) are presented in Section \ref{sec:correlations:results}.

\subsection{Analysis Method}
\label{sec:correlations:method}

\begin{figure}
\resizebox{0.75\columnwidth}{!}{
\includegraphics{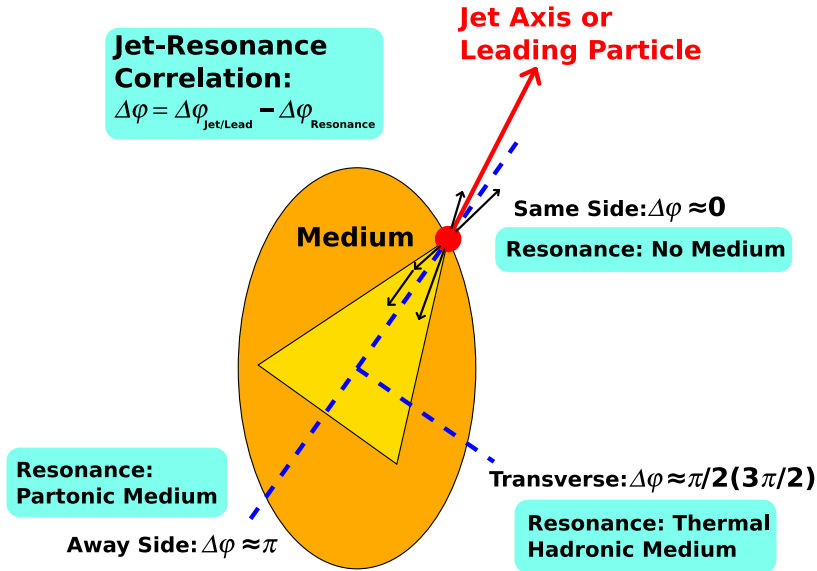} }
\caption{Schematic representation of the jet (or hadron)-resonance correlation method (adapted from \cite{Markert_corr}).  See the text for discussion.}
\label{fig:correlations:diagram}
\end{figure}

The hadron-resonance correlation method is described in detail in \cite{Markert_corr}; only a summary is given here.  The angular correlation function between a jet axis (or a high-$p_{\mathrm{T}}$ trigger hadron acting as a proxy for the jet) is computed.  A heavy-ion collision can be divided into three regions with respect to the jet axis (see Figure \ref{fig:correlations:diagram}).  In the ``near side," where the angular difference $(\Delta\varphi)$ between the trigger axis and the resonance is small, both the resonance and the jet are more likely to have been produced near the surface of the medium.  Therefore any medium modification of the resonance, such as a mass shift or an increase in the width, are expected to be small.  In the region transverse to the trigger axis $(\Delta\varphi\sim\pi/2)$ the resonance signal is expected to be dominated by thermal production in the hadronic medium; modifications to the resonance signal due to medium interactions are again expected to be minimal.  However, on the ``away side" $(\Delta\varphi\sim\pi)$ from the trigger jet or hadron, a resonance is more likely to have interacted with the medium.  It is expected that hadronic reinteraction processes (rescattering and regeneration) will affect the resonance signal for $p_{\mathrm{T}}<$ 2 GeV/$c$.  However, resonances on the away side with $p_{\mathrm{T}}>$ 2 GeV/$c$ are more likely to have interacted with the medium and have decay daughters with enough momentum to escape with minimal interaction in the hadronic phase.  Therefore, the method proposes that the resonances most likely to exhibit the signatures of chiral symmetry restoration are those on the away side from a trigger jet or hadron with $p_{\mathrm{T}}>$ 2 GeV/$c$.

\subsection{Results}
\label{sec:correlations:results}

\begin{figure}
\resizebox{1.0\columnwidth}{!}{
\includegraphics{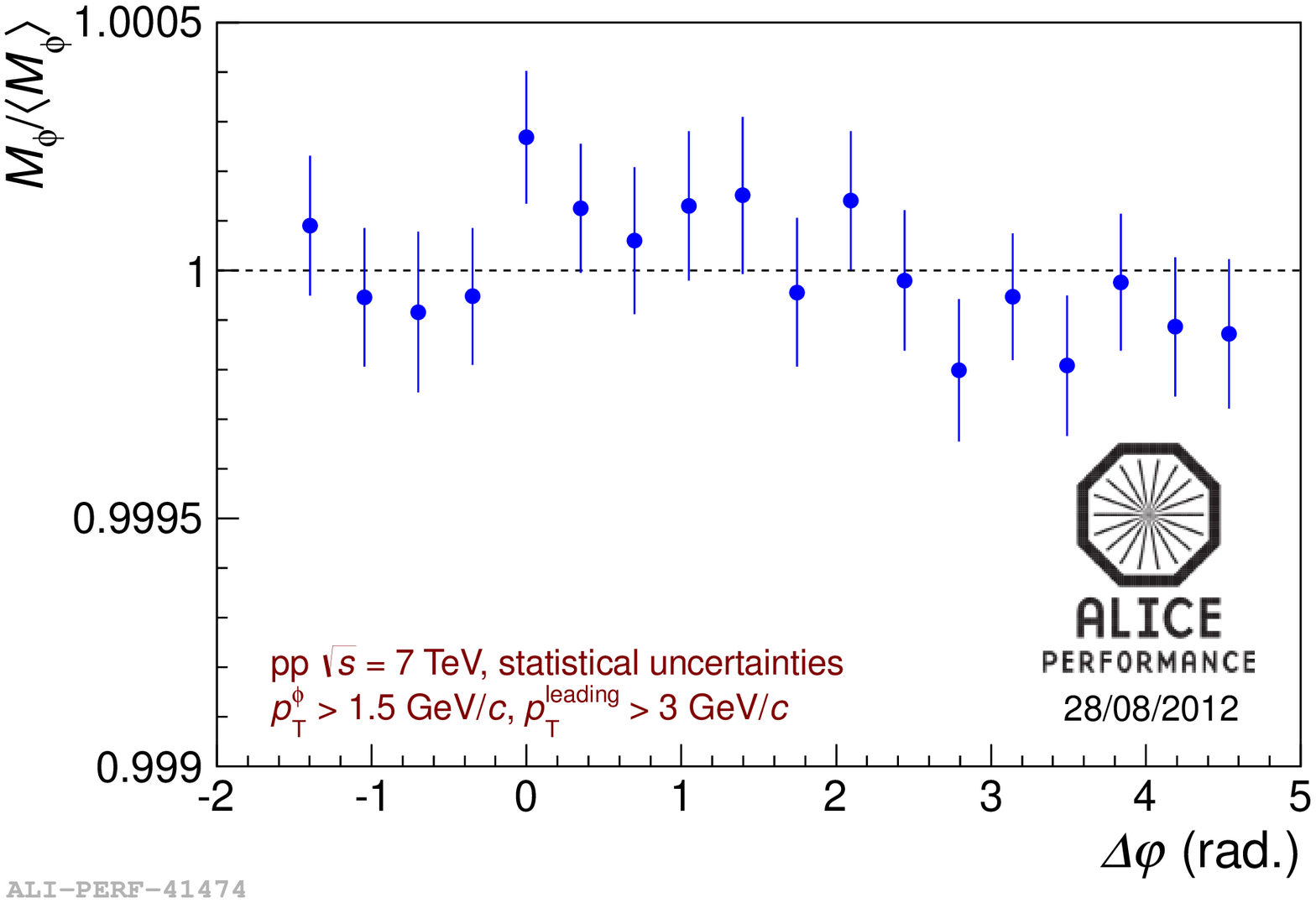}
\includegraphics{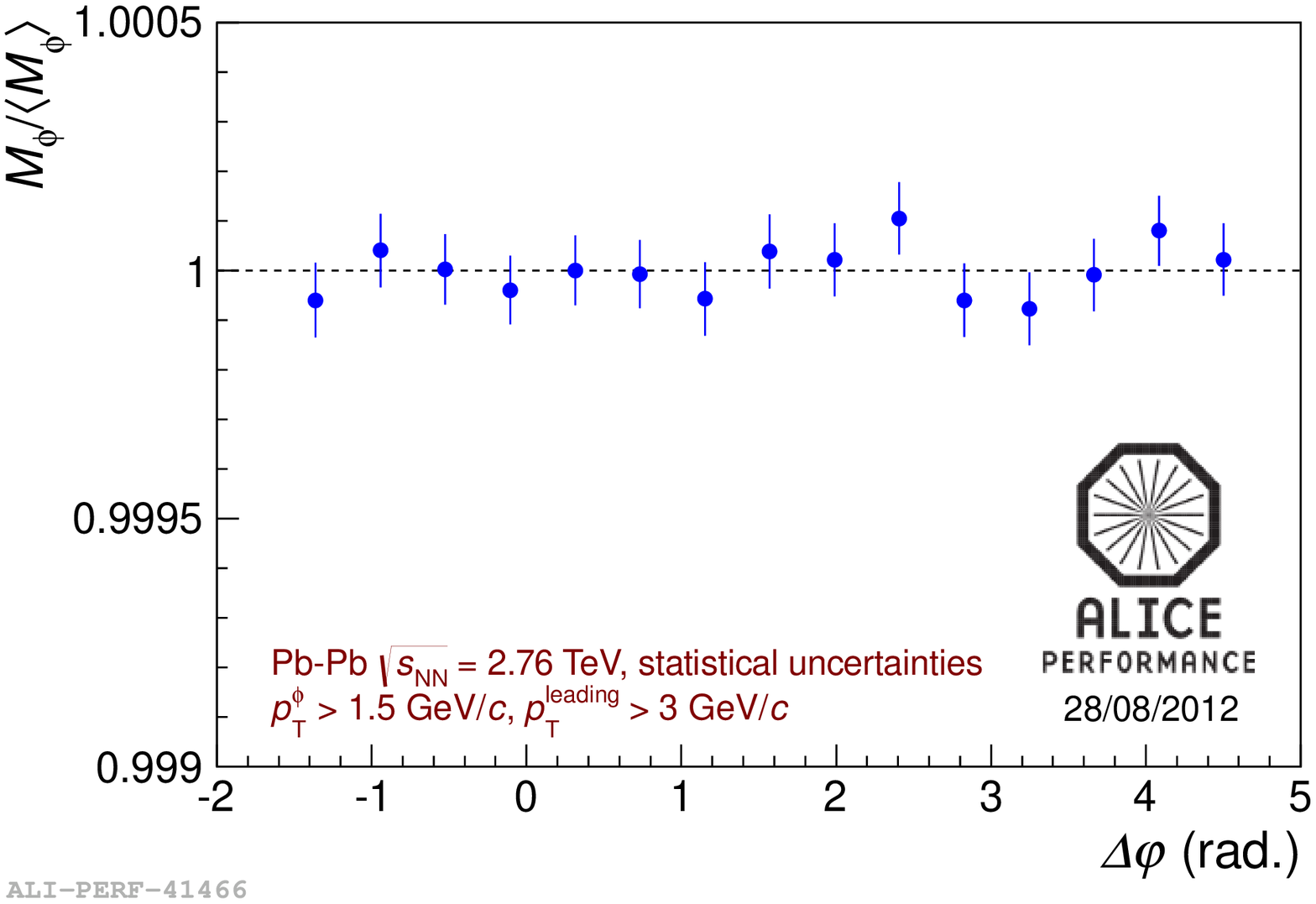} }
\resizebox{1.0\columnwidth}{!}{
\includegraphics{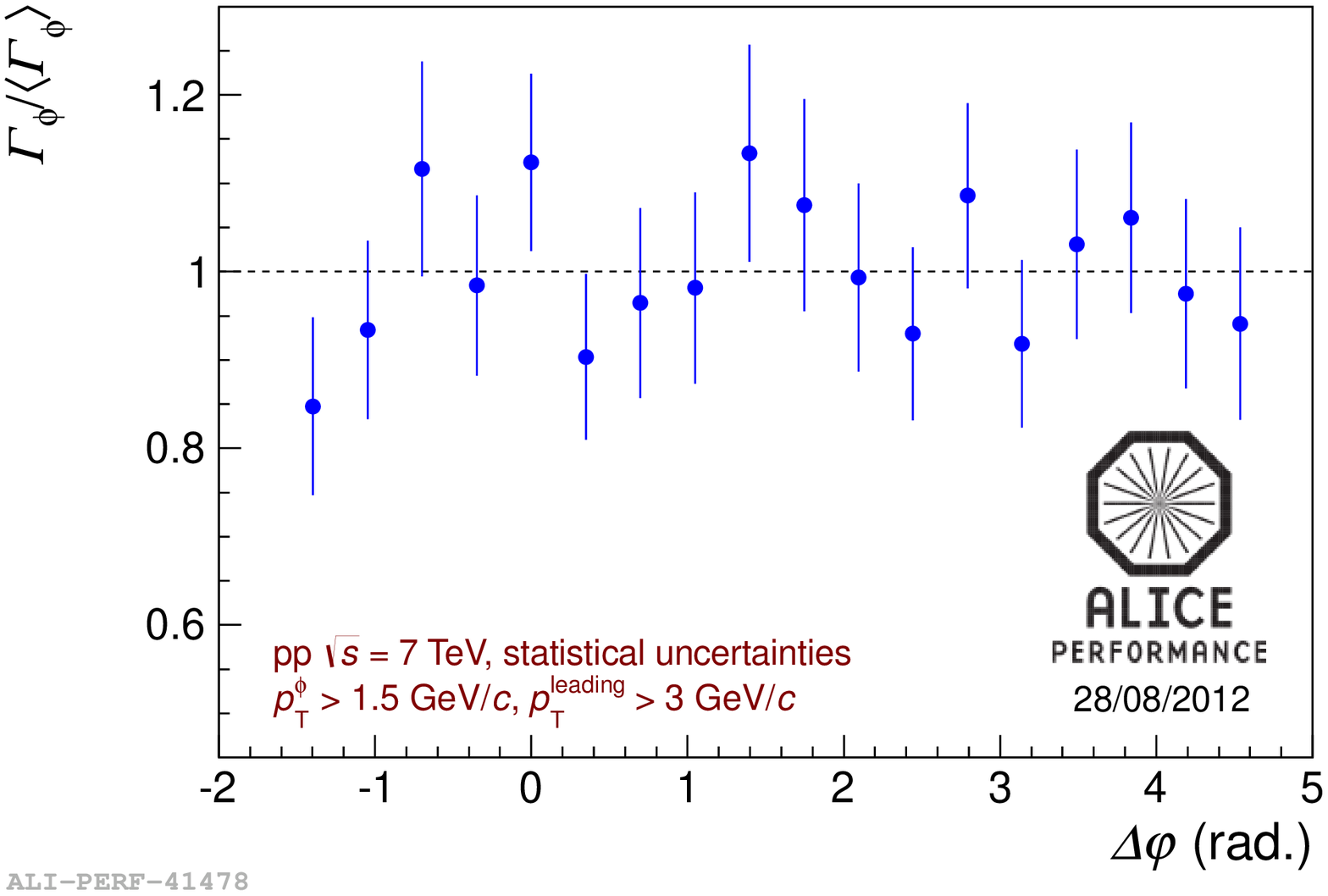}
\includegraphics{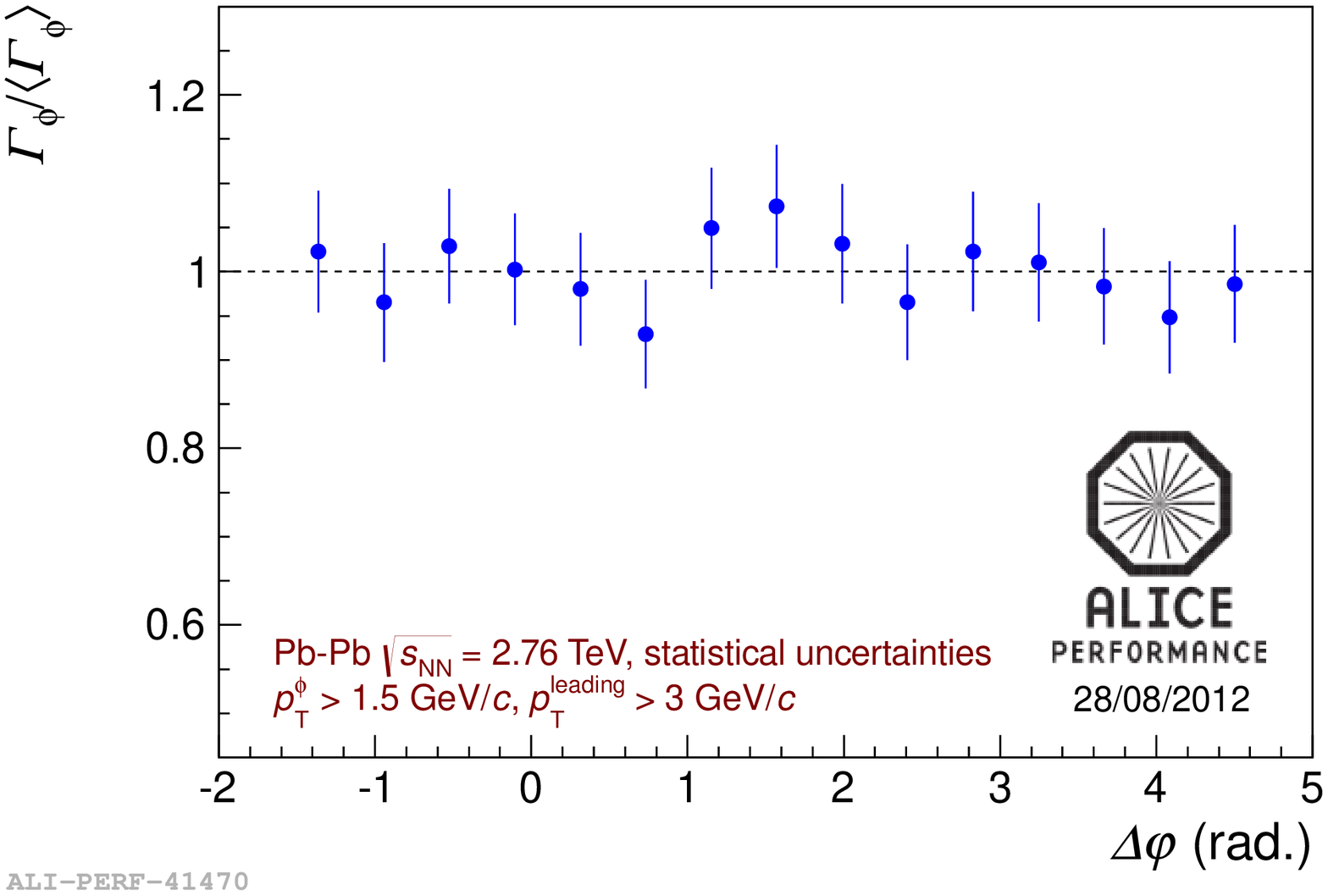} }
\caption{Mass (upper row) and width (lower row) of $\phi$ mesons as functions of correlation angle $\Delta\varphi$ with a leading trigger hadron; pp collisions  at \ssvn on left, Pb--Pb collisions at \stwo on right.  The results in each plot have been divided by their mean value.  Momentum ranges: $p_{\mathrm{T}}^{\phi}>$ 1.5 GeV/$c$, $p_{\mathrm{T}}^{\mathrm{leading}}>$ 3 GeV/$c$.}
\label{fig:correlations:mass_width}
\end{figure}

The difference in azimuth $\Delta\varphi$ between leading trigger hadrons with $p_{\mathrm{T}}^{\mathrm{leading}}>$ 3 GeV/$c$ and $\phi$ mesons has been measured in Pb--Pb collisions at \stwo~\cite{Markert_SQM2011}.  The $\phi$ meson transverse momentum has been restricted to $p_{\mathrm{T}}^{\phi}>$ 1.5 GeV/$c$.  Figure \ref{fig:correlations:mass_width} (upper row) shows the $\phi$ meson mass as a function of $\Delta\varphi$ for pp and Pb--Pb collisions.  The mass measurements for each collision system have been divided by their mean value.  There is no apparent shift in the mass in the away side of the \mbox{Pb--Pb} distribution relative to the near side.  Figure \ref{fig:correlations:mass_width} (lower row) shows the $\phi$ meson width (divided by their mean) as a function $\Delta\varphi$ for pp and Pb--Pb collisions.  No shift in the width is observed in the away side of the \mbox{Pb--Pb} distribution relative to the near side.

\section{Conclusions}
\label{sec:conclusions}

Measurements of the yields of hadronic resonances in heavy-ion collisions and the ratios of those yields to non-resonance yields can be used in conjunction with thermal models to estimate the chemical freeze-out temperature of the system and the time between chemical and thermal freeze-out.  The uncorrected spectra of $\phi$ and $\mathrm{K}^{*0}$ mesons have been extracted by reconstructing the decays $\phi\rightarrow \mathrm{K}^{-}\mathrm{K}^{+}$ and $\mathrm{K}^{*0}\rightarrow\pi^{\pm}\mathrm{K}^{\mp}$.  In the future, the corrected spectra, total yields, and ratios to non-resonances will be calculated.  A shift in the mass or an increase in the width of a resonance in heavy-ion collisions may be a signature of chiral symmetry restoration.  The masses and widths of $\phi$ and $\mathrm{K}^{*0}$ mesons have been measured as functions of transverse momentum for multiple centrality bins.  No centrality dependence is observed in these quantities for either species.  Deviations in the $\phi$ mass and width from their vacuum values have been observed, though similar deviations are observed when simulation data are analyzed.  The $\mathrm{K}^{*0}$ masses measured in Pb--Pb collisions are consistent with the values measured in pp collisions.  The $\mathrm{K}^{*0}$ widths are consistent with the vacuum value.  Resonances with $p_{\mathrm{T}}>$ 2 GeV/$c$ on the away side from a high-$p_{\mathrm{T}}$ jet or trigger hadron may be more likely to exhibit the signatures of chiral symmetry restoration.  Angular correlations of $\phi$ mesons with respect to a leading hadron have been measured; the $\phi$ meson mass and width as functions of $\Delta\varphi$ have been calculated.  Neither a shift in mass nor an increase in width are observed in the aways side of Pb--Pb collisions.  It should be noted, however, that the lower transverse-momentum limits used in this analysis ($p_{\mathrm{T}}^{\phi}>$ 1.5 GeV/$c$ and $p_{\mathrm{T}}^{\mathrm{leading}}>$ 3 GeV/$c$) may be too low and that higher cutoff values may be necessary to sufficiency enrich the sample with resonances that interacted with a chirally-restored medium.


\begin{thebibliography}{}
\bibitem{Bleicher_Aichelin} M. Bleicher and J. Aichelin, Phys. Lett. B, \textbf{530} 81-87
\bibitem{Bleicher_Stoecker} M. Bleicher and H. St\"{o}cker, J. Phys. G: Nucl. Part. Phys., \textbf{30} S111-S118
\bibitem{Markert_thermal} C. Markert \textit{et al.}, Proceedings of PASI 2002 and hep-ph/0206260 (2002)
\bibitem{Vogel_Bleicher} S. Vogel and M. Bleicher, Proceedings of Nucl. Phys. Winter Meeting 2005 in Bormio and nucl-th/0505027v1
\bibitem{Andronic2009} A. Andronic \textit{et al.}, Phys. Lett. B \textbf{673}, (2009) 142-145
\bibitem{PBM2011} P. Braun-Munzinger and J. Stachel, ``Hadron Production in Ultra-relativistic Nuclear Collisions and the QCD Phase Diagram: an Update," \textit{From Nuclei To Stars: Festschrift in Honor of Gerald E Brown}, ed. S. Lee, (World Scientific, Singapore, 2011) and arXiv:1101.3167v1
\bibitem{AndronicQM2011} A. Andronic \textit{et al.}, J. Phys. G: Nucl. Part. Phys., \textbf{38} (2011) 124081
\bibitem{Torrieri_thermal} G. Torrieri and J. Rafelski, Phys. Lett. B \textbf{509}, (2001) 239-245
\bibitem{Karsch} F. Karsch, Nucl. Phys. A \textbf{698}, (2002) 199
\bibitem{Rapp2009} R. Rapp \textit{et al.}, arXiv:0901.3289v1 (2009)
\bibitem{Brodsky_chiral} S. J. Brodsky and G. F. de Teramond, Phys. Rev. Lett. \textbf{60} 1924 (1988)
\bibitem{Holt_Haglin} L. Holt and K. Haglin, J. Phys. G: Nucl. Part. Phys., \textbf{31} (2005) S245-S251
\bibitem{ALICE_detector} K. Aamodt \textit{et al.} (ALICE Collaboration), J. Inst. \textbf{3} (2009) S08002
\bibitem{PDG} J. Beringer \textit{et al.} (Particle Data Group), Phys. Rev. D \textbf{86}, (2012) 010001
\bibitem{ALICE_TPC} J. Alme \textit{et al.}, Nucl. Instr. Meth. in Phys. Res. A \textbf{622} (2010) 316
\bibitem{Markert_corr} C. Markert \textit{et al.}, Phys. Lett. B \textbf{669}, (2008) 92-97
\bibitem{Markert_SQM2011} C. Markert (for the ALICE Collaboration), Acta Physica Polonica B Proceedings Supplement, Vol. 5, No. 2, (2012) 243-248
\end{thebibliography}
\end{document}